\documentclass[fleqn,usenatbib,useAMS]{mnras}

\usepackage{graphicx}	
\usepackage{amsmath}	
\usepackage{amssymb}	
\usepackage{multicol}        
\usepackage{bm}		
\usepackage{pdflscape}	

\usepackage[T1]{fontenc}
\usepackage{ae,aecompl}

\usepackage{newtxtext,newtxmath}

\title[Morphology and Kinematics of HZ7]{An extended [CII] halo around a massive star-forming galaxy at $z=5.3$}

\author[T. S. Lambert et al.]{
Trystan S. Lambert$^{1}$ \thanks{E-mail: TrystanScottLambert@gmail.com},
A. Posses$^{1}$,
M. Aravena$^{1}$, 
J. G\'onzalez-L\'opez$^{2, 1}$, 
R. J. Assef$^{1}$, \newauthor
T. Díaz-Santos$^{1, 3, 4}$,
D. Brisbin$^{1}$,
R. Decarli$^{5}$,
R. Herrera-Camus$^{6}$,
J. Mejía$^{7}$, \newauthor
and C. Ricci$^{1, 8}$ 
\\
$^{1}$   Núcleo de Astronomía, Facultad de Ingeniería y Ciencias,
  Universidad Diego Portales, Av. Ej\'ercito Libertador 441, Santiago, Chile. \\
$^{2}$ Las Campanas Observatory, Carnegie Institution of Washington, Casilla 601, La Serena, Chile \\
$^{3}$ Institute of Astrophysics, Foundation for Research and Technology–Hellas (FORTH), Heraklion, GR-70013, Greece. \\
$^{4}$ Chinese Academy of Sciences South America Center for Astronomy (CASSACA), National Astronomical Observatories, CAS, Beijing 100101, China. \\
$^{5}$ INAF – Osservatorio di Astrofisica e Scienza dello Spazio di Bologna, via Gobetti 93/3, I-40129, Bologna, Italy. \\
$^{6}$ Departamento de Astronomía, Universidad de Concepción, Barrio Universitario, Concepción, Chile.  \\
$^{7}$ European Southern Observatory, Casilla 19001, Santiago 19, Chile\\
$^{8}$ Kavli Institute for Astronomy and Astrophysics, Peking University, Beijing 100871, People’s Republic of China.
}

\date{Last updated 2022 Oct 18; in original form 2022 October 18}

\pubyear{2022}

\begin{document}
\label{firstpage}
\pagerange{\pageref{firstpage}--\pageref{lastpage}}
\maketitle

\begin{abstract}
High-redshift observations are often biased towards massive and bright galaxies that are not necessarily representative of the full population. In order to accurately study galaxy evolution and mass assembly at these redshifts, observations of ``normal'' main sequence galaxies are required. Here we present Atacama Large Millimeter/Submillimeter Array (ALMA) 0.3" resolution observations of the [CII] emission line at 158$\mu$m of HZ7, a main sequence galaxy at $z=5.25$. Comparing to archival rest-frame UV observations taken by the Hubble Space Telescope (HST), we find strong evidence of the existence of extended [CII] emission, which we estimate to be twice the size of the rest-frame UV emission, yielding one of the first high-redshift objects where a clear signature of a [CII] ``Halo'' has been detected to date. For a matched Sérsic profile with n = 1, we measured a [CII] effective radius of $0.50\pm 0.04$" (3.07$\pm 0.25$ kpc) and an average rest-frame UV effective radius of $0.2\pm0.04$" ($1.48\pm0.16$ kpc). The [CII] morphology and kinematics of the system suggest a merging event resulting in a non rotating disk system. This event could be responsible for the extended [CII] emission. Alternatively, some potential obscured emission could also explain the [CII] to UV size ratio. These results contribute to the growing consensus with respect to the existence of extended [CII] emission around galaxies.
\end{abstract}




\section{Introduction}

Studying the kinematics and structures of galaxies at high redshifts is fundamental in developing a full understanding of mass assembly, galaxy evolution, and exploring the epoch of reionization. However, galaxies in the high-redshift regime ($z>4$) have faint ultraviolet (UV)/optical apparent magnitudes due to cosmic dimming which can significantly increase the difficulty of probing meaningful samples with increasing redshift \citep{Ellis2001,Bradley2008,Zheng2012,Linzer2020,Steinhardt2021}. The objects that are identified and studied at these redshifts are often either luminous quasars, massive starbursting galaxies, or both \citep[e.g.][]{Ellis2001}. To obtain a detailed view of the properties of galaxies in the early universe, it is important to obtain and analyze observations of ``normal'' galaxies, i.e. $L^{*}$ or near $L^{*}$. These galaxies tend to dominate the star-formation density of the universe across all redshifts including at the epoch of reionization \citep[$z>6;$][]{Piero2014}. Probing the physical conditions of these galaxies has only been made possible due to large technological improvements over the past two decades with facilities such as the Atacama Large Millimeter/submillimeter Array (ALMA), The Hubble Space Telescope (HST) after the addition of the Wide Field Camera 3 (WFC3), and recently, the James Webb Space Telescope (JWST). 

HST deep field integrations have identified and quantified galaxies' contribution to the cosmic star formation rate (SFR) density out to $z\sim10$ \citep{Ellis2013,Oesch2016,LiverMore2017, Oesch2018,Stefanon2021,Bouwens2022, Bouwens2022Highz,Mauro2022}. However, the physical mechanisms for growth in these galaxies remain unknown. Studying the morphology, and subsequently, the kinematics of these galaxies proves difficult since UV continuum and Lyman $\alpha$ emission are significantly absorbed by dust and by neutral hydrogen in the intergalactic medium (IGM) respectively \citep{Maiolino2015,Carniani2017,Derck2022}. Therefore, different tracers of star-formation and the properties of the inter-stellar Medium (ISM) gas are required. 

ALMA has allowed for high angular resolution observations of high-redshift galaxies \cite[e.g.][]{Carniani2017}. This is typically done by observing the [CII] 158 $\mu$m line, which is one of the brightest emission lines we can observe \citep[reaching up to 1$\%$ of the total bolometric luminosity;][]{DiazSantos2013, DiazSantos2017}. The [CII] emission is a remarkable tracer of the ISM as it is the main line for cooling neutral gas in normal star-forming galaxies \citep{Capak2015,Matthee2019, Schaerer2020, Bouwens2022}; moreover, it is not affected by obscuration as much as UV and Lyman $\alpha$ emission. This makes [CII] an ideal line to probe the morphologies and kinematics of distant sources, particularly at z>4, where it shifts into more transparent (sub)millimeter atmospheric windows.

A pilot ALMA [CII] survey - carried out using an early version of the array- observed ten galaxies at z $\sim$ 5 \citep{Capak2015}. Nine of them were chosen as Lyman-break galaxies (LBGs) in the Cosmic Evolution Survey (COSMOS) field \citep{Scoville2007} with spectroscopic confirmation of their redshifts from the Deep Extragalactic Imaging and Multi-Object Spectrograph (DEIMOS) on the W.M. Keck-II Telescope in Hawaii (for more information on the selection criteria see \cite{Capak2015}. Comparisons between these observations and rest-frame ultra-violet (UV) observations---taken using the Advanced Camera for Surveys (ACS) on HST---show the infrared excess (IRX $\equiv L_{\rm IR}/L_{\rm UV}$) ratio at a given UV-slope $\beta$ to be substantially lower than expected according to current models, as can be seen in Figure 2 of \cite{Capak2015}. This result was later confirmed by \cite{Barisic2017}. This suggests that there is substantial evolution taking place very early on in the universe, at the epoch of reionization, since current models cannot account for the excess in [CII] emission of these sources.

To determine what evolutionary processes might be occurring within galaxies during this epoch, we have taken high angular resolution observations of some of these galaxies to explore their morphological and kinematic properties. In this paper we study the observations of HZ7 in particular, one of the brightest objects in the \cite{Capak2015} sample, with a rest-frame UV spectroscopic redshift of z = 5.2532±0.0004 \citep{Capak2015}. This galaxy seems to be a ``normal'' galaxy based on its location with respect to the main sequence of star forming galaxies (near $L^{*}$), at those redshifts \citep{Capak2015, Speagle2014}. Furthermore, it has good multiwavelength coverage \citep{Leauthaud2007, Capak2015, ANA2022}, which makes it an excellent target.

 Of particular interest is the idea of ``[CII] haloes'', which have been suggested in order to explain the observed distribution of gas and the star-formation in high-redshift galaxies \citep{Fujimoto2019,Fujimoto2020,Ginolfi2020Halo}. Previous studies have identified extended [CII] emission, or [CII] haloes, using low-resolution ALMA imaging \citep[$\sim$ 1.0" beam;][]{Fujimoto2020,Ginolfi2020}. So far, only two similar main sequence (``normal'') galaxies have been studied with high-resolution [CII] imaging (0.3"), in this redshift range, enabling a clear determination of the existence of a [CII] halo: \cite{Herrera-Camus2021} found a [CII] halo whilst \cite{ANA2022} did not. There are several suggested physical mechanisms that might lead to such haloes including the presence of satellite galaxies, circumgalactic H$_{\rm II}$ regions, cold streams of gas, outflows, mergers, and photodissociation (these physical mechanisms are discussed in more detail below).

We note that this is an unusual use of the term ``halo'', as it is often used to describe particular astrophysical components of a galaxy, for example: dark matter haloes, stellar haloes, and galactic haloes---which are usually spheroidal in shape. However, the wide-spread use of the term in the recent literature has provided a strong precedent for [CII] haloes to refer to extended, fuzzy [CII] emission around high-redshift galaxies \citep{Fujimoto2019, Fujimoto2020, Pizzati2020, Ginolfi2020Halo, Herrera-Camus2021, ANA2022}. In order to remain consistent with the current literature on the topic, we will make use of the term ``[CII] halo'' but we emphasize that we strictly refer to extended [CII].

Here we present high signal-to-noise ALMA [CII] follow-up observations of HZ7 at an angular resolution comparable to that of HST and JWST. The new observations allow us to probe of the [CII] and the rest-frame UV emission at similar scales, and search for potential extended [CII] emission.We discuss previous as well as our own observations of HZ7 in $\S 2$. The analysis and results are described in $\S 3$, followed by a discussion in $\S 4$. Our summary and conclusions are found in $\S 5$. Throughout this paper we adopt a vanilla $\Lambda$CDM cosmology with $\Omega_{m} = 0.3$, $\Omega_{\Lambda} = 0.7$, and $H_0 = 70~\rm km~\rm s^{-1}~\rm Mpc^{-1}$.

\section{Data}

\subsection{Target Selection and Physical Properties}
HZ7 is part of the sample of galaxies presented by \cite{Capak2015}. These objects were selected as LBG candidates within the COSMOS field, with spectroscopically determined redshifts. They all have luminosities between 1 and 4 $L^{*}$ and with $\beta$ between -1.4 and -0.7 \citep{Capak2015}.

HZ7 was detected at a redshift of z = 5.25 and was detected by \cite{Capak2015} in [CII] with a peak flux density of $\sim 2$ mJy. An upper limit of 36 $\mu$Jy was determined from the 158$\mu$m continuum emission, implying that there may be low dust content or that HZ7 is an outlier in the IRX-$\beta$ relation. \cite{Capak2015} also estimated a SFR of $21_{-2}^{+5}$ $\rm{M}_{\odot} \rm{yr}^{-1}$ using the \cite{Kennicutt1998} relation \citep{Ota2014, Capak2015} and a combination of UV and far-infrared data, as well as an upper limit of 10.35 L$_{\odot}$ for the IR luminosity. The IR Luminosity was determined by fitting grey body models, available in \cite{Casey2012}. \cite{Capak2015} estimated a stellar mass of 9.86 $\pm 0.21$ M$_{\odot}$ and a dynamical mass of 10.8$^{+1.5}_{-1.0}$ M$_{\odot}$ by fitting Bruzual and Charlot templates \citep{Bruzual2003} to COSMOS photometry as done previously in \cite{Ilbert2013} and using the method in \cite{wang2013} to derive the dynamical mass from the [CII] velocity dispersion.

\subsection{ALMA Observations}
The target HZ7 was observed as part of the ALMA
project 2018.1.01359.S (PI: M. Aravena). At the redshift for HZ7 of z = 5.2532±0.0004 \citep{Capak2015} the [CII] line is centered at 303.9292 GHz which falls into the ALMA band 7 (275 GHz – 373 GHz). The observations were carried out on 2018-11-28 with a total of 80 minutes of on-source integration time.

During the observation, two overlapping spectral windows (SPWs) were placed to detect the [CII] emission and two were placed in the opposite side-band to map the continuum at a rest-frame wavelength of 161.5$\mu$m. The ALMA observatory staff, as part of standard pipeline, performed initial data calibration. The calibrated visibility data were then re-analyzed, performing additional flagging of bad time periods in the data and bad channels. We concatenated the new ALMA observations with that obtained by \cite{Capak2015}. The previous data was taken in a more compact configuration providing extra sensitivity to extended emission. The data were re-imaged using a natural weighting of the u-v visibility data to create continuum, moment zero, and channel maps with an average beam size of 0.27". Gaussian fitting of the spectral and spatial data was performed using the Common Astronomy Software Applications (\textsc{CASA}) software viewer and associated fit tools \citep{CASAPAPER}. The continuum and line images were interactively cleaned by manually masking out the emission.

Since the residual map which results from the CLEAN algorithm can have incorrect default flux scaling when combining different array configurations \citep{JVM1995}, we apply a correction as suggested by \cite{JvMModern2021}. This correction makes an important difference with regards to the total flux density measurements of HZ7 (resulting in a $~33\%$ decrease in the peak flux density), but very little, to no qualitative difference with respect to the spatial distribution. We remark that the main results presented in this paper remain the same when using only the new ALMA data and no artificial structures are added when combining with data from Capak et al. (2015).

\subsection{HST Observations}
HST/WFC3 observations of HZ7 in the F105W (10430.83	\AA), F125W (12363.55 \AA), and F160W (15278.47 \AA) bands were obtained from the Hubble Legacy Archive (PID:13601), and are used in this work to compare the rest-frame UV emission as traced by these images with the [CII] emission observed by ALMA. These data products were corrected astrometrically by calibrating them to the GAIA DR2 release \citep{GAIADR2}. We queried the DR2 release using the python package \texttt{astroquery.gaia} \footnote{\url{https://astroquery.readthedocs.io/en/latest/gaia/gaia.html}}. Using the positional information within the HST headers and then manually associated each GAIA star with a star in the HST image. The difference in position between the DR2 release and the actual stars in the HST images renders the offset which is then applied to the HST images. The average offset was $\sim$ 0.300 $\pm$ 0.008" ($\Delta \alpha = -0.128"$; $\Delta \delta = 0.283"$) \footnote{Full GAIA calibration and routine is available at \url{https://github.com/TrystanScottLambert/Hz7_ISM/blob/main/Align.py}}. 
\section{Analysis and Results}

\subsection{Spatially integrated emission line}

\begin{figure}
    \includegraphics[width=0.5\textwidth]{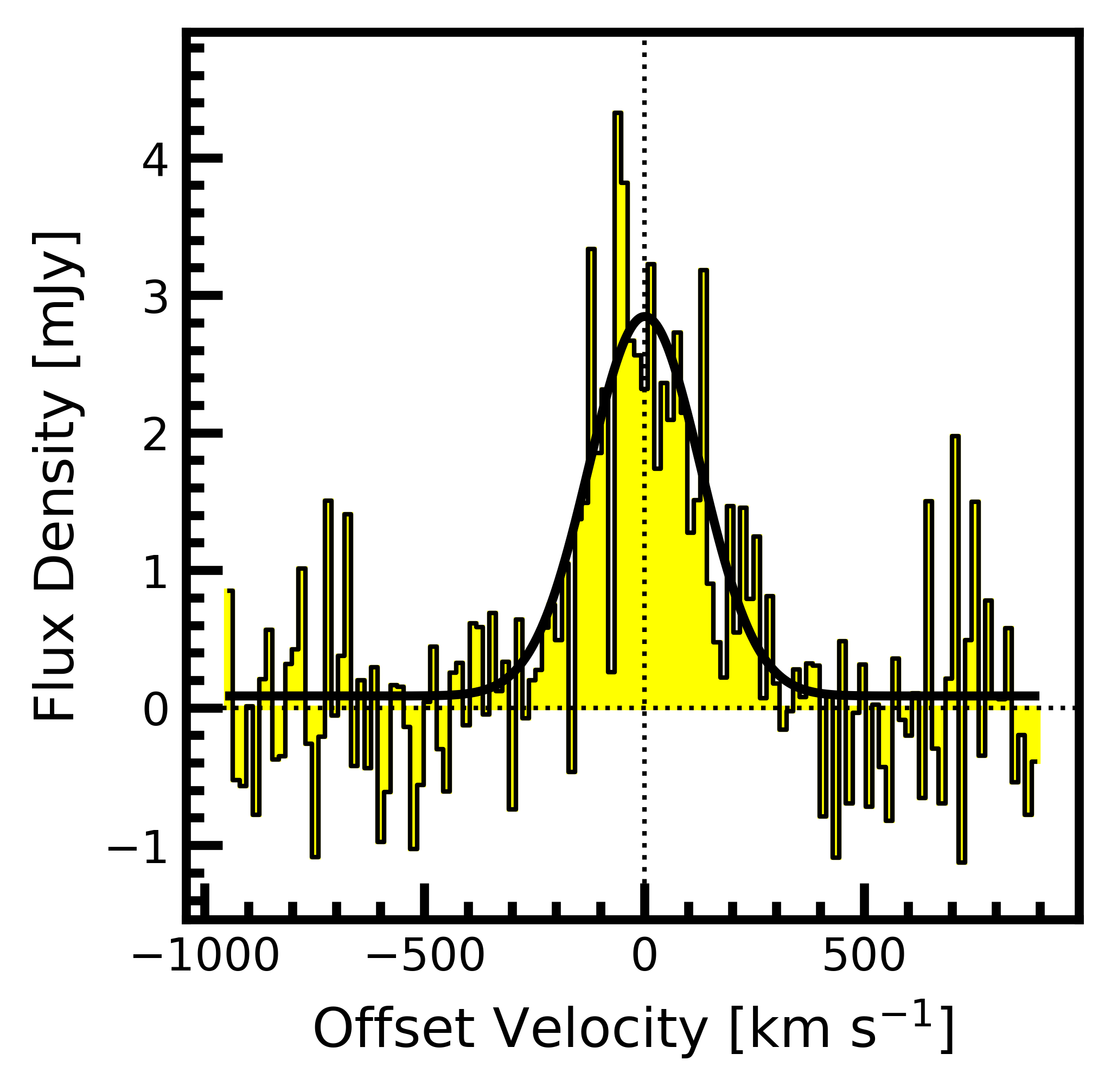}
    \caption{Spatially integrated spectrum of [CII] emission toward HZ7 at z = 5.25, at a channel (spectral) resolution of 15 km s$^{-1}$, using a combination of the new high resolution data and those reported in Capak et al. (2015).}
    \label{fig:IntegratedSpectrum}
\end{figure}

Figure \ref{fig:IntegratedSpectrum} shows the spatially integrated spectrum of HZ7. This plot was created following the procedure in \cite{Bethermin2020}, which was also used in \cite{Endsley2022} and \cite{Schouws2022}. This is an iterative process where a small aperture equal to the area of the beam width is centred on the main source of emission ($\alpha$ = 09:59:30.459, $\delta$ = 2.08.02.586); using this aperture, an integrated spectrum is extracted and a Gaussian profile can be fitted to the emission; the full width at half maximum (FWHM) from this fit is used to create a simple moment-0 map which in turn provides a mask---by taking all emission greater than 3$\sigma$---and then masks the entire cube. This masked cube is used to extract another integrated spectrum, and the process is repeated, until the extracted integrated spectrum remains unchanged. The shape and integrated flux of the spectrum is similar with that obtained by \cite{Capak2015} with a FWHM of $298 \pm 27 $~km s$^{-1}$ and an integrated flux density ($S_{\rm [CII]}$) of $0.86~\pm~0.11$~Jy~km~s$^{-1}$ (as compared to $ 380 \pm 42$ km s$^{-1}$ and 0.71 $\pm$ 0.07 Jy km s$^{-1}$). We calculate the [CII] line luminosity to be $\log\left(L_{\rm [CII]}/L_{\odot}\right)~=~8.8 \pm 0.1$---very close to the value of 8.74, estimated by \cite{Capak2015}.


\begin{table}
\caption{Measured and derived properties as well as literature values of Hz7.}
\begin{tabular}{lll}
\hline
Property               & Value & Reference \\ \hline \hline
$z_{\rm [CII]}$        & $5.2532 \pm 0.0004$     & \cite{Capak2011}         \\
$S_{ \rm [CII]}$ &  $0.86 \pm 0.11$ Jy km s$^{-1}$   & this work    \\ 
log$\left(L_{\rm [CII]}\right / L_{\odot})$      & $8.8 \pm 0.1$     & this work         \\
{[}CII{]} FWHM         & $298 \pm 27$ km s$^{-1}$    & this work         \\
$\sigma_{\rm [CII]}$ & $127 \pm 12$ km s$^{-1}$ & this work \\
RA                     & 149.876986$^{\circ}$      & \cite{Capak2015}         \\
Dec                    & 2.134113$^{\circ}$     & \cite{Capak2015}         \\
SFR                    & $21_{-2}^{+5}$ M$_{\odot}$ yr$^{-1}$    & \cite{Capak2015}         \\
$Y_{\rm 105W}$ &     $24.56 \pm 0.04$  &   \cite{Barisic2017}         \\ 
$J_{\rm 125W}$ &     $ 24.37 \pm 0.04$  &   \cite{Barisic2017}         \\ 
$H_{\rm 160W}$ &     $24.28 \pm 0.04$  &   \cite{Barisic2017}         \\ 
$\log\left({M_{\odot}}\right)$& $9.86 \pm 0.21$     & \cite{Capak2015}         \\ \hline \hline
\end{tabular}
\label{tbl:properties}
\end{table}

\subsection{Spatial and velocity distribution of [CII]}

\begin{figure}
    \centerline{\includegraphics[width= 0.52\textwidth]{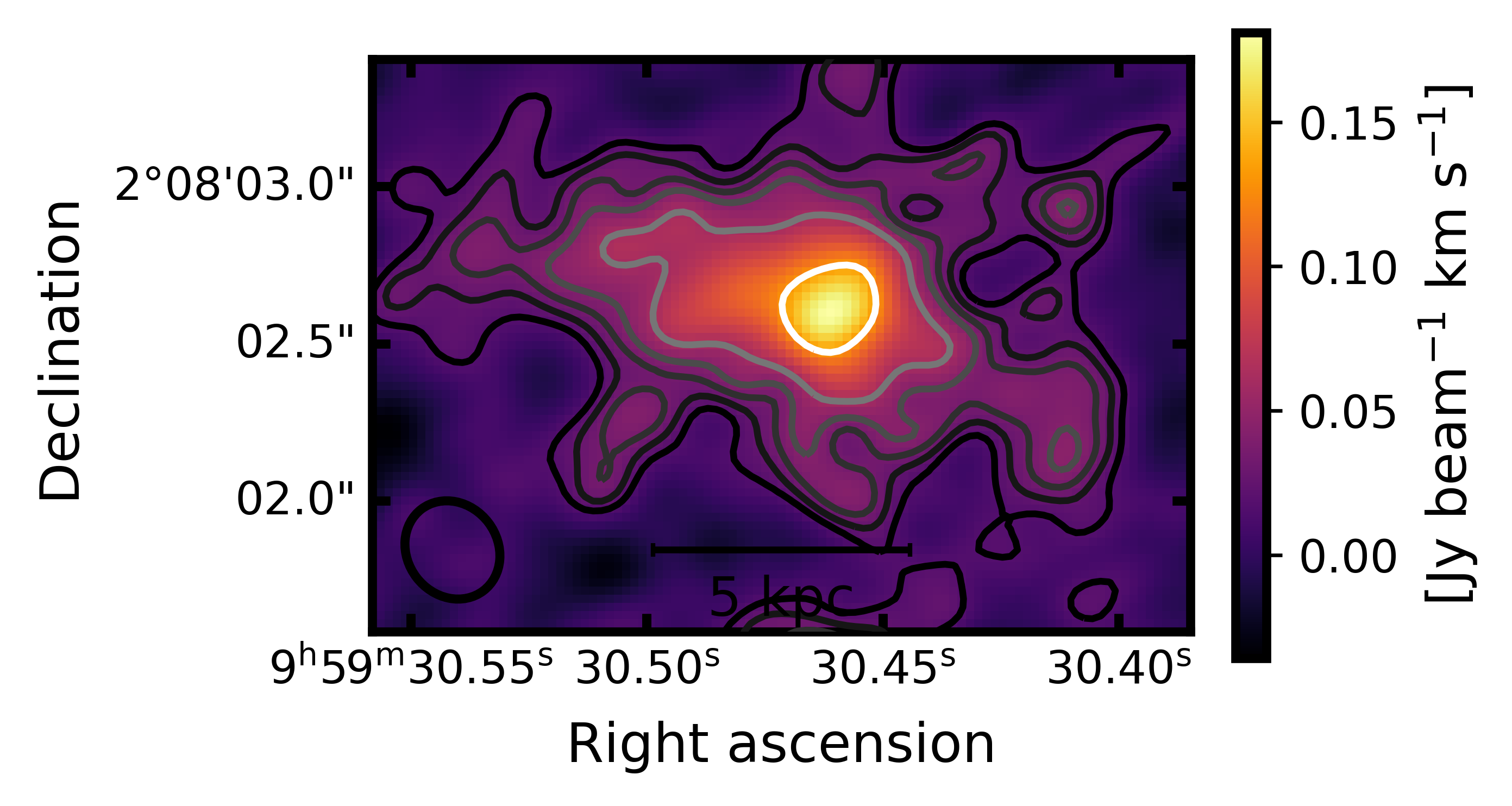}}
    \centerline{\includegraphics[width= 0.52\textwidth]{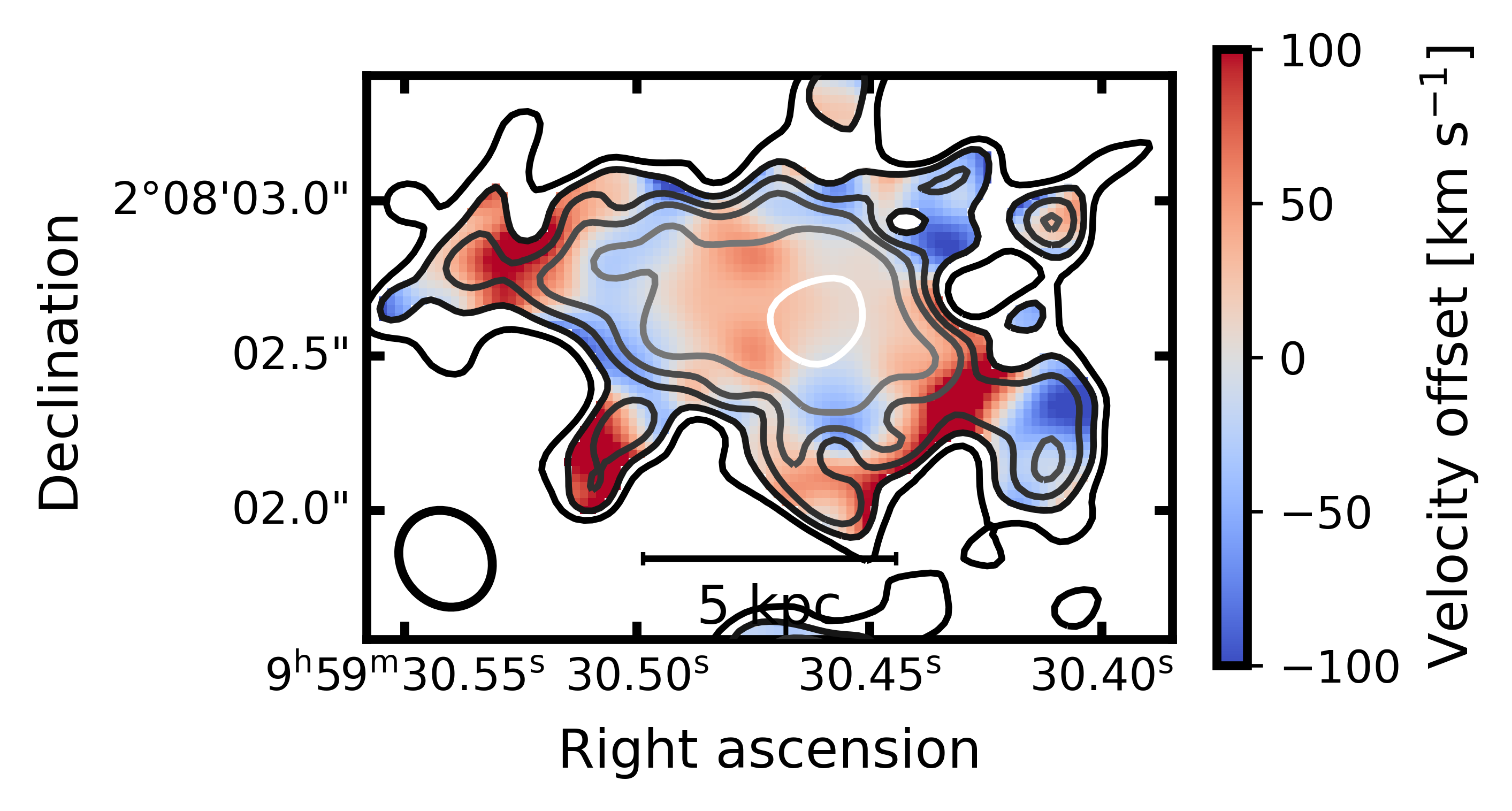}}
    \centerline{\includegraphics[width= 0.52\textwidth]{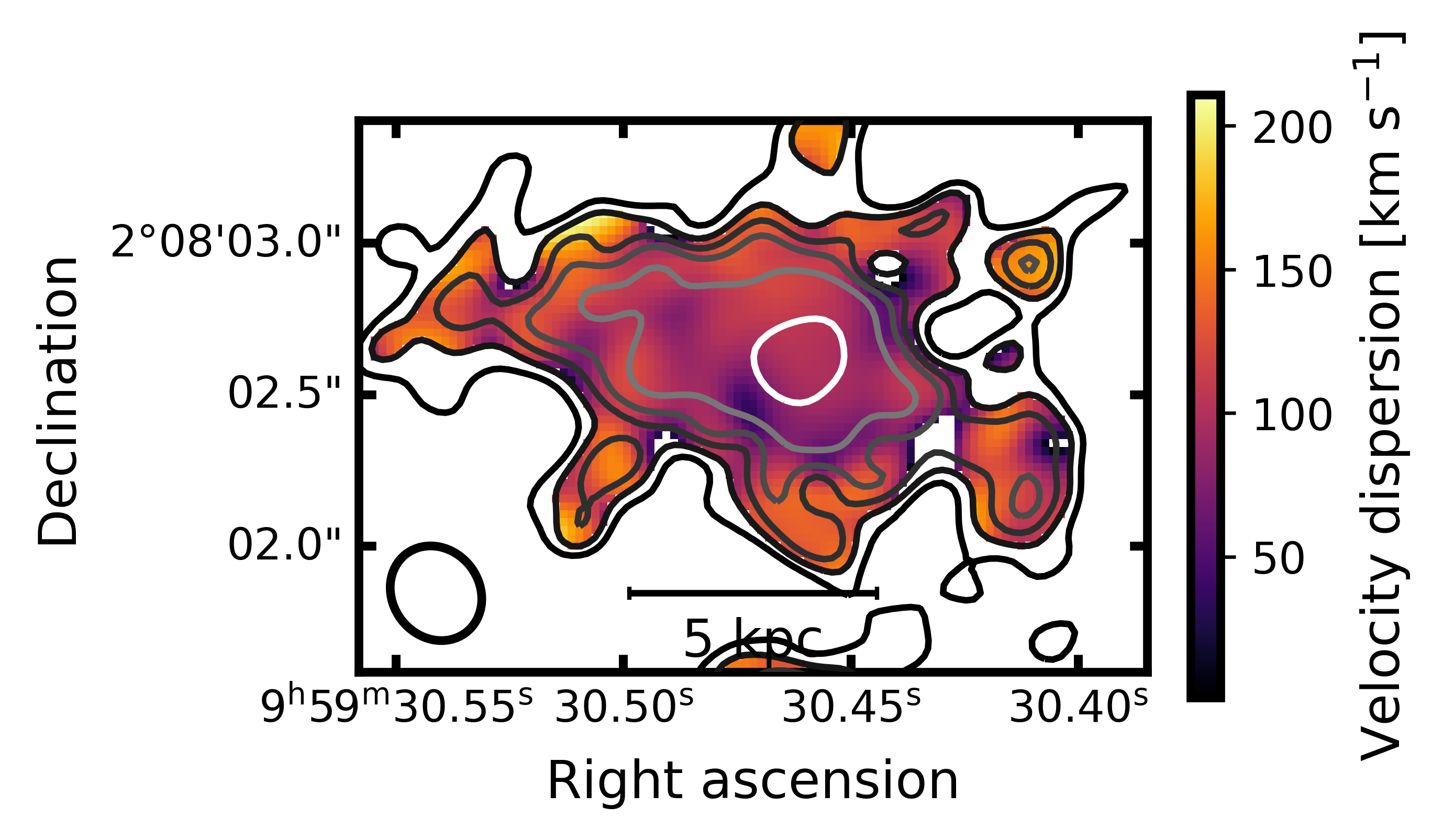}}
    \caption{TOP: Integrated [CII] emission the HZ7 galaxy with $\sigma = 0.02$ Jy beam$^{-1}$ km s$^{-1}$. Contours show the 2, 3, 4, 5, 7, and 15 $\sigma$ levels of the rms of the integrated emission. These contours are overlaid in the moment-1 and moment-2 maps below. MIDDLE: [CII] Velocity Field of the HZ7 galaxy. BOTTOM:   Dispersion map of the [CII] detection of the HZ7 galaxy.}
    \label{fig:moments}
\end{figure}

The frequency integrated [CII] emission, velocity field, and dispersion map are shown in Figure \ref{fig:moments}. The channels which were collapsed to create these maps, were selected by using the Gaussian fit in Figure \ref{fig:IntegratedSpectrum}. Specifically, we chose all channels within the FWHM, which resulted in 18 channels on either side of the central emission channel---a velocity range of 540 km s$^{-1}$. 

The higher-order moment maps provide essential information but can often be difficult to interpret without masking; i.e. focusing only on the area of the map which contains the emission and ignoring the rest. We used the integrated [CII] emission to build masks for the velocity field and dispersion map, by taking all emission greater than 3$\sigma$ within the moment-0 map. The integrated emission of HZ7 shows a galaxy with a centrally bright concentration of light. However, the emission is elongated in the NE direction with significant structure. This can be seen by the tendril-like appendage to the north east as well as the small knots of emission around the center of the galaxy.
 
Contrary to the integrated [CII] emission, the velocity field and dispersion map do not show ordered structures. Both higher-order moment maps are disordered with the velocity field in the middle panel of Figure \ref{fig:moments} showing no indication of a gradient and therefore no clear indication of a rotating disk. Likewise, the velocity dispersion map in the bottom panel of Figure \ref{fig:moments} shows significant structure which is in stark contrast to the well-behaved brightness distribution of the integrated [CII] emission. The disturbed morphology, and the likely turbulent ISM---inferred from the large emission line width---suggests that HZ7 could possibly be a merging system and not a single galaxy. 

Both the integrated spectrum (Fig. \ref{fig:IntegratedSpectrum}) and the integrated [CII] emission (top panel of Fig. \ref{fig:moments}) are used to derive several source properties of HZ7 in the following sections. The source information can be found in in Table \ref{tbl:properties}, along with previous measurements for this source in the literature.

\subsection{Spatial distribution of [CII], rest-frame UV, and dust continuum}
\begin{figure}
    \centering
    \includegraphics[width=0.5\textwidth]{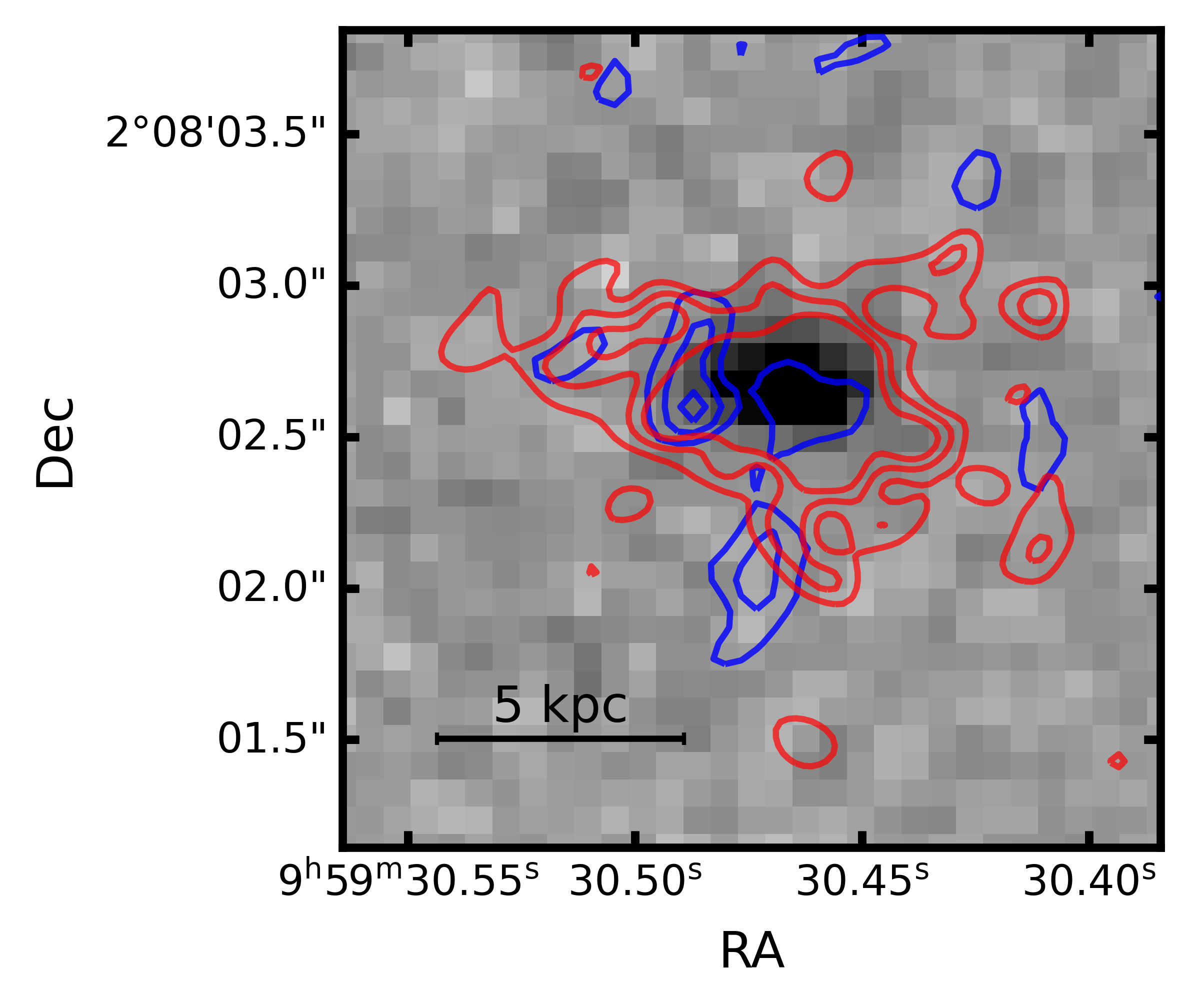}
    \caption{Comparison of the rest-frame UV, [CII] and dust continuum emission toward HZ7. The gray-scale background represents the Y band HST observation. Red contours (3-5 $\sigma$) represent the integrated emission from the ALMA [CII] data whilst the blue shaded contours (2-5 $\sigma$) are that of the continuum emission.}
    \label{fig:Continuum}
\end{figure}

We present the spatial distribution of the dust continuum emission in Figure \ref{fig:Continuum} with blue contours. The continuum dust emission, unlike the [CII] emission, is not smooth and is instead clumpy with a central source overlapping the central emission in both [CII] (contoured in red), and the rest-frame UV (represented by the background in gray-scale). Figure \ref{fig:Continuum} shows that the spatial distribution of the [CII] emission is extended beyond that of the rest-frame UV as well as the continuum. The dust continuum emission appears to be more extended than the rest-frame UV and contains a southern extension which surpasses even the [CII] emission. More detailed discussion on the physical interpretation of Figure \ref{fig:Continuum} follows in \S 4.

\subsection{[CII] compared to rest-frame UV}
\begin{figure*}
    \centering
    \includegraphics[width=\textwidth]{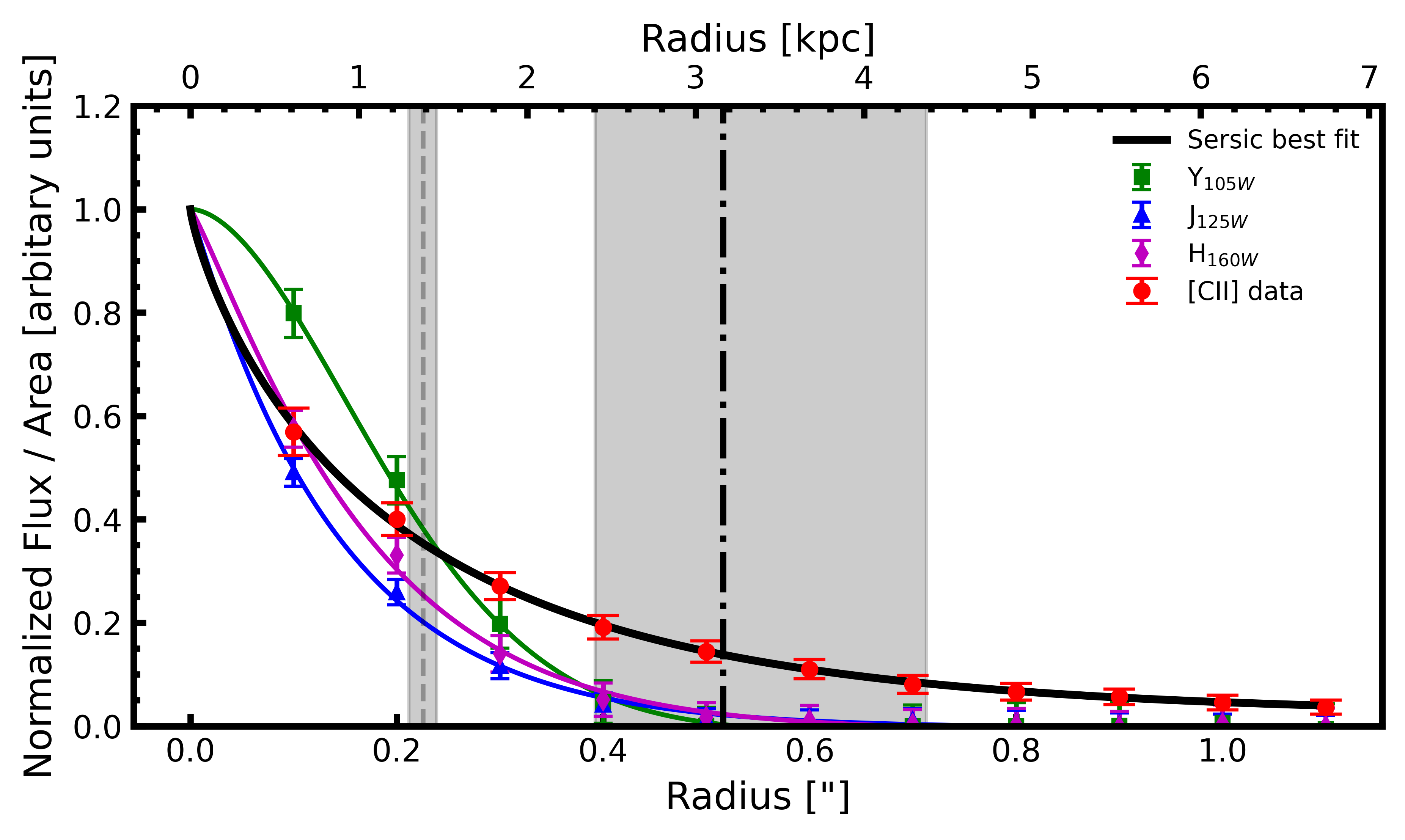}
    \caption{Fitted S\'ersic profiles of the [CII] with a Sérsic index of 0.9 (black line) as well as the UV emission from HST in the F105W, F125W, F160W and total UV emission (shown as green, blue, and magenta lines respectively). The vertical dashed and dot-dashed lines highlight the average effective radius from the UV Sérsic fits and the effective radius from the [CII] Sérsic fits respectively.}
    \label{fig:RadialProfiles}
\end{figure*}

A radial profile was generated by placing a series of increasingly larger annuli, in bins of 0.1", around the peak [CII] emission of the moment-0 map. This peak was determined by summing up the emission along the RA and Dec axis independently and then determining the central RA and Dec values which were calculated from the resulting two, independent, 1-D Gaussian fits. The total flux in each annulus was then calculated and normalized by the area represented by each respective annulus. The uncertainty of each annulus was calculated as 
\begin{equation}
    u\left(F\right) = \sqrt{n\sigma^{2} + F},
\end{equation}
for the HST images and
\begin{equation}
    u\left(F\right) = \sigma \sqrt{\frac{n}{n_{\rm beam}}},
\end{equation}
for the [CII] moment-0 map, where $\sigma$ is the RMS, $F$ is the total flux within the annulus, $n$ is the number of pixels within an annulus, and $n_{\rm beam}$ is the number of pixels in a beam area.
The results of constructing this radial profile can be seen in Figure \ref{fig:RadialProfiles}. A Sérsic profile was adopted in order to fit the radial profiles using a Monte-Carlo Markov-chain fitting routine, with the python package \texttt{emcee} \cite{emcee}.

To appropriately compare morphologies between the HST and ALMA data, the high resolution HST data was convolved with the ALMA beam. The Sérsic free-index radial profiles of all the available HST data---F105W, F160W, and F125W---are compared to the free index Sérsic fit of the [CII] emission and is shown in Figure \ref{fig:RadialProfiles}. The centers of the individual HST filter images were determined in the same manner as the [CII] profile (explained earlier); therefore, any residual systematic offset between the [CII] and rest-frame UV emission do not affect the final results. The best-fit effective radii and Sérsic indices are given in Table \ref{tbl:revalues}. The comparison between the [CII] and rest-frame UV radial profiles show that the effective radius of the [CII] emission is 2 - 3 times larger than that of the UV emission. The two dashed lines show the $R_{\rm e}$ parameter of the S\'ersic fit of the [CII] emission and the average $R_{\rm e}$ parameter for all the HST filters. To compare all the fitted profiles in the same scale, and to be comparable to what is usually adopted in the literature, we also used a fixed Sérsic index of 1.0. In this case, the obtained [CII] sizes are still 2x larger than those of the rest-UV. 

In addition to the UV and [CII] radial profiles, we also compare their surface distributions in Figure \ref{fig:Continuum}. Although more affected by S/N, the surface distribution analysis allows the added benefit of spatial information which itself can be important to the understanding of the underlying physical processes. We once again see that the the [CII] emission is far more extended than that of the rest-frame UV emission. This is qualitative confirmation of this extension found for the radial profiles.


\begin{table}
\caption{Best-fit parameters for the Sérsic fits of the HST filters as well as the [CII] emission.}
\begin{tabular}{llll}
\hline \hline
Wavelength     & $R_{\rm e}$ ["]& $R_{\rm e}$ [kpc] & Sérsic Index \\ \hline\hline
[CII] & $0.59^{+0.08}_{-0.10}$ & $3.60^{+0.49}_{-0.64}$ & $1.28^{+0.21}_{-0.27}$\\\\
Y$_{105W}$ & $0.20^{+0.02}_{-0.03}$ & $1.23^{+0.11}_{-0.16}$ & $0.55^{+0.13}_{-0.22}$\\\\
J$_{125W}$ & $0.24^{+0.04}_{-0.11}$ & $1.47^{+0.24}_{-0.69}$ & $1.04^{+0.37}_{-0.87}$\\\\
H$_{160W}$ & $0.23^{+0.04}_{-0.10}$ & $1.41^{+0.21}_{-0.58}$ & $0.83^{+0.29}_{-0.74}$\\
\hline
[CII] & $0.50^{+0.04}_{-0.04}$ & $3.07^{+0.23}_{-0.25}$ & 1\\\\
Y$_{105W}$ & $0.24^{+0.02}_{-0.03}$ & $1.48^{+0.15}_{-0.16}$ & 1\\\\
J$_{125W}$ & $0.24^{+0.02}_{-0.03}$ & $1.44^{+0.14}_{-0.15}$ & 1\\\\
H$_{160W}$ & $0.23^{+0.03}_{-0.03}$ & $1.51^{+0.17}_{-0.18}$ & 1\\
\hline
\hline
\end{tabular}
\label{tbl:revalues}
\end{table}

\section{Discussion}
Recent studies of massive star forming galaxies at z=4-6 have found the almost ubiquitous presence of [CII] line emission extending beyond the location of star forming regions traced by the rest-frame UV \citep{Fujimoto2019,Fujimoto2020}. Such [CII] observations have been mostly conducted at low angular resolution (1.0") or through stacking analysis, hindering the physical mechanisms that might produce such haloes or making it unclear whether the extended [CII] emission is partly due to the nature of the observations. HZ7 is amongst the first galaxies at these redshifts ($z>4$) with clear evidence for a smooth and wide [CII] emission, along with HZ4 \citep{Herrera-Camus2021}.

\subsection{Morphology and Kinematics}

HZ7 has a series of clumpy [CII] regions. The integrated emission shown in Figure \ref{fig:moments} shows knotted regions of significant [CII] concentration, in addition to the main central [CII] emission. There are potentially several physical reasons that could explain these isolated peaks. However, no counterparts to these structures can be found in the rest-UV emission map shown in Figure \ref{fig:Continuum}. Figure \ref{fig:moments} shows that the main, central [CII] emission has a ``smooth'' distribution. The central [CII] emission and the central UV emission are well aligned but the dust continuum is less present within this central region and has a reasonably large southern extension without any UV emission. 

The velocity map (middle panel of Fig. \ref{fig:moments}) does not have any gradient and does not show any preferred state of motion. The possible explanations are varied and might include dynamical disturbance. This would be evidence of a possible merger \citep{ALPINE, Romano2021}. This is further backed by the dispersion map (bottom panel of Fig. \ref{fig:moments}) which, once again, does not show any kind of preferred, or ordered motion. Both the moment-1 and moment-2 maps show that this is a complex kinematical system.

\subsection{[CII] Halo}
In \S 3.4 we presented evidence of a large [CII] emission which is roughly double or triple that of the UV emission. A possible interpretation of such a result is the existence of a [CII] halo \citep{Fujimoto2019,Ginolfi2020Halo,Pizzati2020}. What makes the case of HZ7 interesting is that unlike \cite{Fujimoto2019} and \cite{Ginolfi2020Halo}, our results are not derived through stacking of a large sample of galaxies, but are rather inferred from direct measurements of the halo extension in a single target. \cite{Fujimoto2019} stacked a sample of 18 normal galaxies (including HZ7) in the redshift range $5.153 \leq z \leq 7.142$. Both \cite{Fujimoto2019} and this work's UV results fall well within the average values expected for normal star-forming galaxies at these redshifts as calculated by \cite{Shibuya2015}. Notably, we found the ratio of the [CII] effective radius to the rest-frame UV effective radius ($R_{\rm eff, [CII]} / R_{\rm eff, UV}$) to be $\sim 3$ which strongly agrees with the result from \citet[see their Fig. 3]{Fujimoto2020} and exceeds the ratio found by \cite{Herrera-Camus2021} of 1.2, who detected a [CII] halo within HZ4, which occupies a similar position to HZ7 on the main sequence.

\cite{Fujimoto2020} provides a test to determine if a system has a [CII] halo: masking out a beam-sized area of central [CII] emission, and all emission greater than 10 kpc, which results in an annulus of [CII]; if the remaining emission is greater than 4$\sigma$, then the system can be considered to have a [CII] halo. HZ7 meets this requirement. However, as has been stated by \cite{Herrera-Camus2021}, this test is purely observational in nature; it is only dependent on the galaxy size vs the beam area and does not take into account any other information besides the [CII] observation, ignoring rest-frame UV data. Therefore, \cite{Herrera-Camus2021} suggest a more physical, but similar, definition: measuring if the extended [CII] emission---the emission which extends further than the rest-frame UV emission---is greater than $4\sigma$. In both definitions, HZ7 is considered a system with a [CII] halo.

\subsubsection{Physical Interpretations}
\cite{Fujimoto2019} lists 5 physical possibilities for [CII] haloes in high-redshift galaxies: (i) the existence of satellite galaxies, (ii) circumgalactic photodissociation regions, (iii) circumgalactic H$_{\text{II}}$ regions, (vi) cold streams, and (v) outflows. In the case of HZ7, it is not possible to determine the exact physical process or processes behind the observed [CII] halo from these ALMA observations alone. \cite{Fujimoto2019} could only go so far as to exclude satellite galaxies as the physical reason due to a lack of stellar continuum emission within the haloes but could not definitively differentiate amongst the other four. Later, \cite{Pizzati2020} proposed a physical model to explain these [CII] haloes---suggesting that supernovae-driven cooling outflows are the most likely reason \citep[see][]{Ginolfi2020Halo}.

\begin{figure*}
    \centering
    \includegraphics[width = \textwidth]{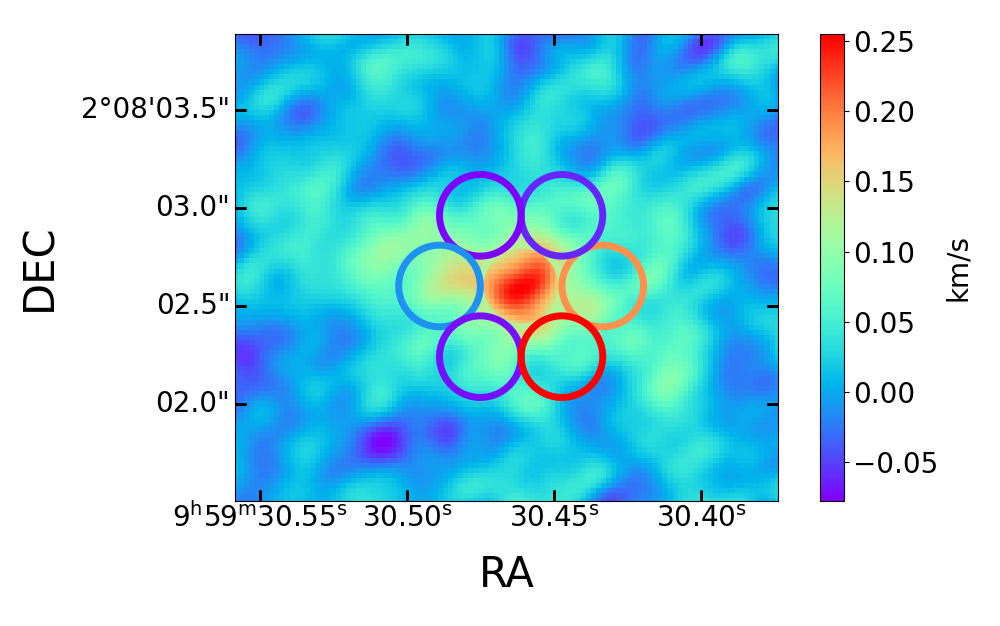}
    \caption{Circular Analysis results as per Herrera-Camus et al. (2017). Each circle represents an aperture that was used to extract a spectrum along the velocity axis. These spectra were fitted with gaussians-fits. The resulting central velocity was used to color each respective circle. }
    \label{fig:circular analysis}
\end{figure*}

We checked for extended velocity components across the galaxy, which would imply the presence of an outflow, following the analysis of \cite{Herrera-Camus2021}. This involved placing seven small apertures (once with radii of 0.05" and once with 0.02") around the source (Fig. \ref{fig:circular analysis}). The [CII] profiles of each of these apertures were then examined. However, we found that there was no evidence for an outflow, which would be identified as an offset or wider secondary line components in the spectra, at the sensitivity of our observations. \cite{Herrera-Camus2021} managed to measure these offsets in HZ4 thus suggesting outflows as the physical process associated with the creation of that halo. This could imply that either deeper, or higher signal-to-noise, observations might be necessary in order to detect evidence of outflows. 

Hints about the origin of this halo can be taken from the continuum emission shown in Figure \ref{fig:Continuum}. Both the dust continuum and the [CII] have some emission overlapping the peak of the UV emission. The eastern emission in continuum overlaps with the eastern extension of [CII], but is lacking in UV emission. Detectable dust emission in this area combined with a discernible lack of UV emission suggests that the latter is being suppressed due to dust obscuration. This could possibly explain the existence of [CII] haloes in the first place, implying that what we observe as [CII] haloes are in fact due to dust obscuration suppressing the total UV emission. But based on these results of HZ7, the galaxy would be dust poor towards the center, which seems unlikely. There would also be an expectation for the [CII] and dust emission to be aligned, and dust emission towards the outskirts of the system. Both of these criteria are not observed in this data. In particular, current theory suggests that galaxies tend to have dust centrally concentrated \citep{Gomez2010, Mosenkov2019}, which makes this explanation even more unlikely.

As suggested by \cite{Fujimoto2019}, it is possible that [CII] haloes are produced by multiple clumps which could be either in-falling satellites or outflowing gas. However, we only find an extended diffuse emission, with no signs of bright clumps, at the resolution of our observations (0.3" $\sim $ 2 kpc). This implies that if indeed the extended emission is produced by clumps, they would have to be numerous and significantly smaller than 2 kpc.


A point of particular interest in the continuum emission in Figure \ref{fig:Continuum}, is the presence of a Southern continuum extension, extending well beyond the bounds of the [CII] emission. This could possibly correspond to an interloper, which would result in the [CII] emission being redshifted out of the band, or alternatively this dust emission might be a merger remnant which would imply that HZ7 is a late stage merger. This would be in agreement with the disturbed velocity field and dispersion map in Figure \ref{fig:moments} as well as the -15 km s$^{-1}$ panel in Figure \ref{fig:ChannelMaps} in the Appendix. However, while both of these potential scenarios would be in agreement with the existence of this southern continuum emission, they are both unlikely as the probabilities of their occurrences are rare. It is entirely possible that the continuum data does not have a high enough SNR in order to make any firm deductions. Higher S/N observations of the continuum would be needed in order to fully investigate its contribution

Another explanation for the extend [CII] emission is gas stripping due to tidal interactions which would exist in a merging system. In this case, the [CII] would be excited by shocks, as has been observed at low redshifts by \cite{appleton2013}, in Stephan's Quintet. This explanation is supported by the ratio between the [CII] and FIR emission in the system. \cite{Barisic2017} calculates an upper FIR limit of $\log\left(L_{\rm FIR}/L_{\odot}\right) < 10.35$, resulting in a lower limit ratio $\log\left(L_{\rm [CII]}/L_{\rm FIR}\right)>1.5$, which would imply shock-enhanced [CII] emission according to \cite{appleton2013}. A similar physical argument is made in \cite{Ginolfi2020Halo}, at higher redshifts ($z\sim4.57$). This physical explanation of the existence of a [CII] halo further supports the possibility that HZ7 is a merging system.


 
\section{Summary and Conclusions}
HZ7 is a so-called ``normal'' galaxy at $z=5.25$ for which the physical properties (morphological and kinematical) are still not well understood. We have presented new high resolution [CII] line emission observations with ALMA of HZ7 with the aim of exploring its kinematics and morphology. 

HZ7 is unequivocally not a simple rotating disk galaxy and this has been shown by the [CII] velocity field and the [CII] dispersion map (middle and bottom panels of Figure \ref{fig:moments}). The overall structure of both the moment-1 and moment-2 maps suggest that HZ7 could possibly be a merger. This is further supported by the southern continuum emission in Figure \ref{fig:Continuum}. Determining whether or not the system is actually a merger would require further inspection and study, including follow-up high resolution ALMA observations. 

We have also presented strong evidence of the existence of a [CII] halo which surrounds the UV emission. Whilst the existence of such haloes has been suggested \citep{Fujimoto2019,Ginolfi2020Halo}, this is amongst the first clear, high resolution, direct observational evidence. The [CII] emission has an effective radius of 3.60$^{+0.49}_{-0.64}$ kpc, for a free Sérsic index (and an effective radius of $3.07^{+0.23}_{-0.25}$ for a fixed Sérsic index of 1) whilst the average UV emission amongst the three filters $\sim 1.4$ kpc---indicating a [CII] halo extending 3.3 times the UV emission.

This result provides theoretical models with another example in order to explain the physical process which might result in the existence of such a halo, as contributing to the growing number of case studies of [CII] haloes around ``normal'' galaxies at these redshifts. Which could imply an important and essential role in the galactic evolution of ordinary galaxies. 

Follow-up observations of this system as well as other similar systems will be essential in determining the evolutionary aspects of ordinary galaxies in epoch of reionization. One such project which will help place this particular galaxy in context of other ``normal'' galaxies is ALMA-CRISTAL survey \footnote{\url{https://www.alma-cristal.info/home}} which is a cycle 8 ALMA program which will observe 19 similar galaxies in [CII].

\section*{Acknowledgements}
We thank the anonymous referee for the valuable comments.
This paper makes use of the following ALMA data: ADS/JAO.ALMA$\#$2018.1.01359.S. ALMA is a partnership of ESO (representing its member states), NSF (USA) and NINS (Japan), together with NRC (Canada), MOST and ASIAA (Taiwan), and KASI (Republic of Korea), in cooperation with the Republic of Chile. The Joint ALMA Observatory is operated by ESO, AUI/NRAO and NAOJ. The National Radio Astronomy Observatory is a facility of the National Science Foundation operated under cooperative agreement by Associated Universities, Inc.

MA acknowledges support from FONDECYT grant 1211951 and CONICYT+PCI+REDES 190194. TSL, MA, RJA, and AP acknowledge support from grant CONICYT + PCI + INSTITUTO MAX PLANCK DE ASTRONOMIA MPG190030. MA, RJA, and CR acknowledge support from ANID BASAL project FB210003.
RJA and CR were further supported by FONDECYT grant number 1191124 and by FONDECYT grant number 11190831 respectively.

\section*{Data availability}
The datasets were derived from sources in the public domain: HST images from \url{https://hla.stsci.edu/hlaview.html}; ALMA data from \url{https://almascience.nrao.edu/aq/?result_view=observation}. The data underlying this article will be shared on reasonable request to the corresponding author.

\bibliographystyle{mnras}
\bibliography{trystansbib}

\section*{Appendix}

\begin{figure*}
    \centering
    \centerline{\includegraphics[width=\textwidth]{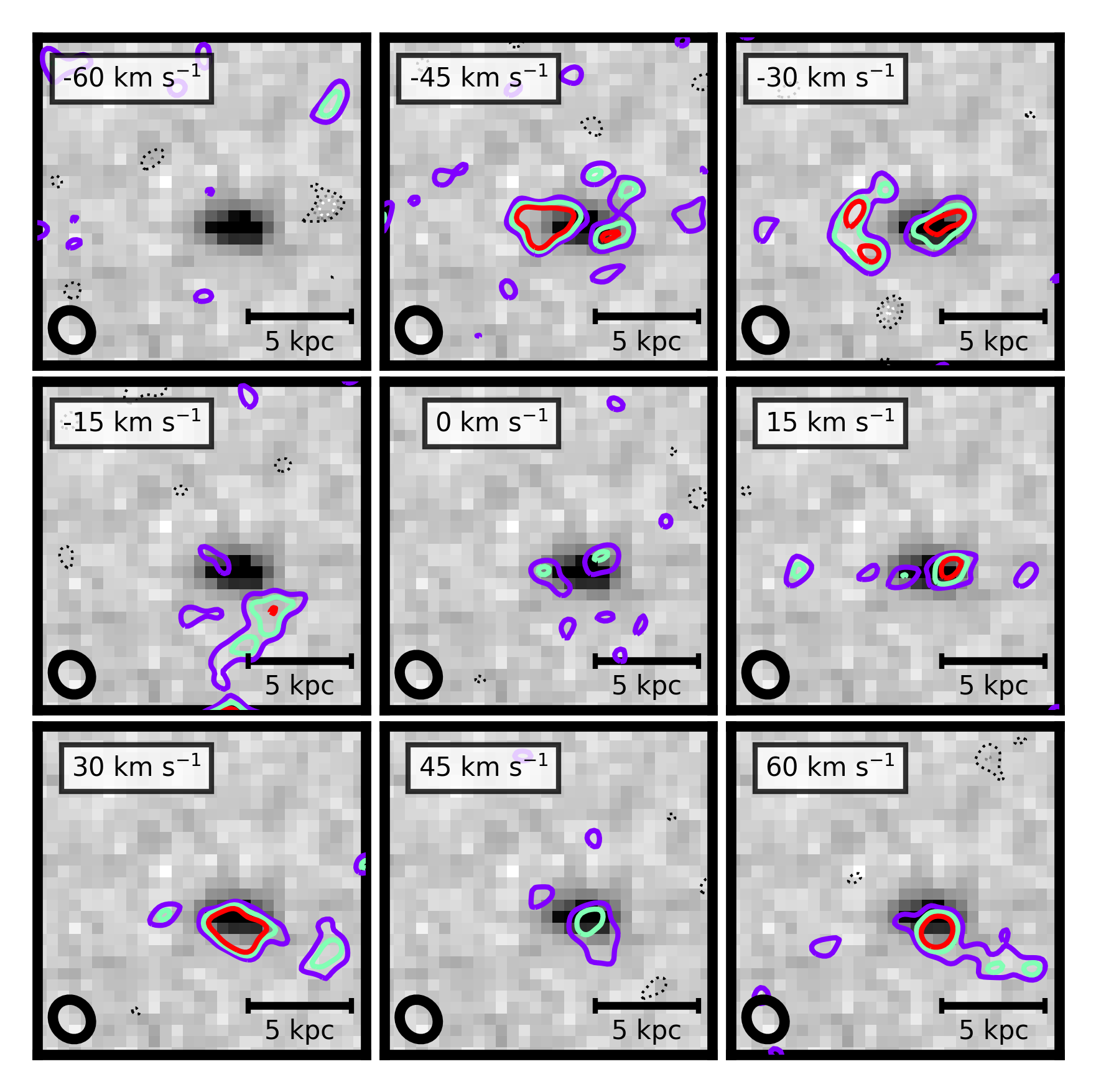}}
    \caption{Channel Map for $\pm$ 60 km s$^{-1}$ around the central emission.}
    \label{fig:ChannelMaps}
\end{figure*}

\bsp	
\label{lastpage}
\end{document}